# Nonlinear heat conduction in Coulomb-blockaded quantum dots

Miguel A. Sierra[a] and David Sánchez[a]

[a]*Instituto de Física Interdisciplinar y Sistemas Complejos IFISC (UIB-CSIC), E-07122 Palma de Mallorca, Spain*

**Abstract**

We analyze the heat current flowing across interacting quantum dots within the Coulomb blockade regime. Power can be generated by either voltage or temperature biases. In the former case, we find nonlinear contributions to the Peltier effect that are dominated by conventional Joule heating for sufficiently high voltages. In the latter case, the differential thermal conductance shows maxima or minima depending on the energy level position. Furthermore, we discuss departures from the Kelvin-Onsager reciprocity relation beyond linear response.

## 1. Introduction

Recent advances suggest that nanostructures will play a pivotal role in improving today's thermoelectric efficiencies [1]. The Seebeck coefficient has been measured in quantum point contacts [2], quantum dots [3] and ballistic microjunctions [4]. The results from these experiments can be understood by using linear-response models. However, nanoscale conductors can also become excellent platforms for the investigation and manipulation of electron dynamics far from equilibrium since large voltage and temperature differences can be applied in a controlled way [5,6]. As a consequence, nonlinear effects are clearly visible in the thermovoltage generated in response to a temperature shift. Charging effects are important in these systems. Although a scattering approach to nonlinear thermoelectric effects accounts for temperatures and voltage dependences of the transmission function [7], Coulomb blockade effects fall beyond the scope of this theory.

We have recently considered in Ref. [8] nonlinear thermoelectric effects within a single-impurity Anderson model with constant energy $U$. Our results agree with the experiments of Staring *et al.* [5] and Svensson *et al.* [6]. Additionally, we discussed asymmetries shown in the heat dissipation when the voltage is reversed in a given terminal or when the created power is compared between two leads. In both cases, higher order effects start to dominate the heat current and a change of sign for the rectification factor is found. This striking result demonstrates that a more careful analysis of the nonlinear heat current in Coulomb-blockaded quantum dots is needed. This is the goal of the present work.

Coulomb blockade is not only detected in quantum dots. This universal effect governs the low temperature physics of many nanoscale systems such as molecular bridges [9], carbon nanotubes [10] and optical lattices [11]. Coulomb blockade becomes apparent when the energy scale associated to electron-electron interactions is larger than any other energy in the system (temperature, level broadening). As a result, electrons tunneling from nearby reservoirs into the sample must pay an extra energy that blocks the transport at low voltages. The effect is clearly seen in the conductance plots as a function of both the bias and the gate voltages.

Since electrons carry energy during tunneling, Coulomb blockade also affects heat conduction. Now, the heat current can be driven by both electric fields and temperature biases. The former leads to the Peltier effect, which leads to a reversible heat flow for small voltages in isothermal conditions. When voltages become larger, nonlinearities can give rise to rectification effects. A study of this phenomenon in the heat current is necessary for an accurate assessment of the cooling properties of nanostructured thermoelectric modules [1,12]. In fact, nonlinear contributions have been observed in atomic-scale junctions [13] and interesting predictions have been made for quantum point contacts and nanowires [14,15,16], doped bulk semiconductors [17] and quantum dots with weak interactions [18].

When heat flow is induced by temperature differences, the Fourier law is obeyed if the thermal gradient is small. Larger temperature biases may give rise to rectification effects, a subject that has lately attracted much attention [19-23]. In the following, we will consider Coulomb blockade effects also for temperature-driven heat currents.

## 2. Model

Our investigation is based on the Anderson model where a quantum dot with a charging energy $U$ plays the role of a magnetic impurity coupled to fermionic reservoirs. The total Hamiltonian in expressed as

$$\mathcal{H} = \mathcal{H}_{\text{leads}} + \mathcal{H}_{\text{dot}} + \mathcal{H}_{\text{tun}} \qquad (1)$$

The first contribution corresponds to the Hamiltonian of the leads,

$$\mathcal{H}_{\text{leads}} = \sum_{\alpha k \sigma} \varepsilon_{\alpha k} C^\dagger_{\alpha k \sigma} C_{\alpha k \sigma} \qquad (2)$$

where $\alpha$ labels the left ($\alpha=L$) or right ($\alpha=R$) reservoir. Furthermore, $k$ is the continuous wavenumber, $\sigma$ is the electron spin index and $\varepsilon_{\alpha k}$ is the energy spectrum, which we take as spin-independent (nonmagnetic leads).

In Eq. (1), the dot Hamiltonian reads

$$\mathcal{H}_{dot} = \sum_\sigma \varepsilon_d d^\dagger_\sigma d_\sigma + U d^\dagger_\uparrow d_\uparrow d^\dagger_\downarrow d_\downarrow \qquad (3)$$

with $\varepsilon_d$ the quasilocalized level of the dot. Here we only consider one single level for definiteness. Importantly, $\varepsilon_d$ can be tuned with an external gate voltage.

The hybridization term that establishes the tunnel coupling between localized and continuum states is

$$\mathcal{H}_{tun} = \sum_{\alpha k \sigma}(V_{\alpha k} C^\dagger_{\alpha k \sigma} d_\sigma + V^*_{\alpha k} d^\dagger_\sigma C_{\alpha k \sigma}), \qquad (4)$$

Here, $V_{\alpha k}$ describe tunneling amplitudes.

Once we have defined the Hamiltonian, the electronic current $I$ and the heat flux $J$ can be obtained with the aid of leads' operators. The former is given by the time evolution of the reservoir occupation $n_\alpha = \sum_{k\sigma} C^\dagger_{\alpha k \sigma} C_{\alpha k \sigma}$, whereas the heat current depends on the change rate of the reservoir energy $\mathcal{H}^\alpha_{\text{leads}} = \sum_{k\sigma} \varepsilon_{\alpha k} C^\dagger_{\alpha k \sigma} C_{\alpha k \sigma}$,

$$I_\alpha = -e \frac{d\langle n_\alpha \rangle}{dt} \qquad (5)$$

$$J_\alpha = \frac{d\langle \mathcal{H}^\alpha_{\text{leads}}\rangle}{dt} - \frac{\mu_\alpha}{e} I_\alpha \qquad (6)$$

As is customary, the heat current results from the energy flux measured with respect to the convective term $\frac{\mu_\alpha}{e} I_\alpha$, where $\mu_\alpha = E_F + eV_\alpha$ is the electrochemical potential of reservoir $\alpha$ which consists of two contributions, the common Fermi energy $E_F$ (chemical part) and the applied voltage shift (electric part). For static fields, both $n_\alpha$ and $\mathcal{H}^\alpha_{\text{leads}}$ commute with the total Hamiltonian. As a consequence, in the steady state we find the conservation law for the electrical current $I_L + I_R = 0$, which allows us to define the net current $I \equiv I_L = -I_R$. On the other hand, the heat currents satisfy $J_L + J_R = -IV$ with $V = V_L - V_R$. In what follows, we show calculations for $J \equiv J_L = -J_R - IV$.

Within the nonequilibrium Keldysh-Green's function formalism [24], the charge and heat fluxes become

$$I = \frac{e}{\pi\hbar}\text{Re}\int_{-\infty}^{\infty} dE\, [\sum_{k\sigma} V_{\alpha k} G^<_{\sigma,\alpha k\sigma}(E)] \quad (7)$$

$$J = \frac{1}{\pi\hbar}\text{Re}\int_{-\infty}^{\infty} dE\, [\sum_{k\sigma} \varepsilon_{\alpha k\sigma} V_{\alpha k} G^<_{\sigma,\alpha k\sigma}(E)] - \frac{\mu_L}{e} I \quad (8)$$

$G^<_{\sigma,\alpha k\sigma}(E)$ is the Fourier transform of the lesser Green's function $G^<_{\sigma,\alpha k\sigma}(t,t') = \frac{i}{\hbar}\langle C^\dagger_{\alpha k\sigma}(t')d_\sigma(t)\rangle$. Applying the Langreth rules and after a little algebra, the currents are recast as

$$I = -\frac{e}{\pi\hbar}\int dE\, \sum_\sigma \frac{\Gamma_L \Gamma_R}{\Gamma}\, \text{Im}\, G^r_{\sigma,\sigma}(E)[f_L(E)-f_R(E)] \quad (9)$$

$$J = -\frac{1}{\pi\hbar}\int dE\, \sum_\sigma \frac{\Gamma_L \Gamma_R}{\Gamma}(E-\mu_L)\text{Im}\, G^r_{\sigma,\sigma}(E)[f_L(E)-f_R(E)] \quad (10)$$

where $G^r_{\sigma,\sigma}(E)$ is the dot retarded Green's function, from which the local density of states can be derived.

In Eqs. (9) and (10), we define the level broadening due to interaction with reservoir $\alpha$, $\Gamma_\alpha = 2\pi\rho_\alpha(E)|V_\alpha(E)|^2$. It is proportional to the lead density of states, $\rho_\alpha(E) = \sum_k \delta(E-\varepsilon_{\alpha k})$, and the tunneling probability $|V_\alpha(E)|^2$. We consider the wide band limit and assume that $\Gamma_\alpha$ is independent of the energy $E$. Additionally, $I$ and $J$ in Eqs. (9) and (10) depend on the difference of Fermi functions at the leads, $f_\alpha(E) = 1/[1 + \exp(E-\mu_\alpha)/(k_B T_\alpha)]$, where $\mu_\alpha$ is the electrochemical potential defined above and $T_\alpha = T + \theta_\alpha$ with $T$ the background temperature and $\theta_\alpha$ the temperature shift applied to lead $\alpha$. We remark that phonons can directly contribute to the heat flux (and indirectly to the charge current) but we henceforth consider low temperatures and neglect phonon effects.

The retarded Green's function in Eqs. (9) and (10) can be found using the equation-of-motion technique [24]. This approach yields coupled equations of motion for correlation functions. The resulting (infinite) hierarchy of equations is simplified based on physical grounds. We focus on the regime where electrons interact strongly in the dot (Coulomb blockade) and take into account tunneling processes up to leading order (sequential tunneling). As a consequence, we disregard cotunneling processes (higher-order tunneling via virtual events) and Kondo correlations (many-body spin-flip interactions in the dot), which can be important for strongly coupled dots or for very low temperatures ($T$ smaller than the level broadening). Therefore, we neglect the following higher order correlators: $\langle\langle d^\dagger_{\bar\sigma} C_{\alpha kq} d_\sigma, d^\dagger_\sigma\rangle\rangle$ and $\langle\langle C^\dagger_{\alpha k\bar\sigma} d_{\bar\sigma} d_\sigma, d^\dagger_\sigma\rangle\rangle$, which correspond to virtual charge excitations in the dot, and $\langle\langle C_{\alpha k\sigma} C^\dagger_{\beta k\bar\sigma} d_\sigma, d^\dagger_\sigma\rangle\rangle$, which describes spin excitations. Taking into account these simplifications, the retarded Green function acquires an analytical form,

$$G^r_{\sigma,\sigma}(E) = \frac{1-\langle n_{\bar\sigma}\rangle}{E-\varepsilon_d+i\Gamma/2} + \frac{\langle n_{\bar\sigma}\rangle}{E-\varepsilon_d-U+i\Gamma/2} \quad (11)$$

in terms of the expected value of the dot occupation

$$\langle n_\sigma \rangle = \frac{1}{2\pi i}\int dE\, G^<_{\sigma,\sigma}(E) \quad (12)$$

where the lesser Green function is determined from the Keldysh equation $G^< = i[\Gamma_L f_L(E) + \Gamma_R f_R(E)]|G^r|^2$.

Equation (11) shows that the spectral function $(-\frac{1}{\pi})\text{Im}\, G^r_{\sigma,\sigma}(E)$ has two Coulomb-blockade peaks centered at $E = \varepsilon_d$ and $E = \varepsilon_d + U$, and broadened by the total linewidth $\Gamma = \Gamma_L + \Gamma_R$. Each peak is weighted by the (electron-like or hole-like) occupation for the opposite spin. We stress that the Coulomb blockade regime is only possible when $\Gamma < U$ since both peaks must be visible. We emphasize two features of the transmission function. First, the resonance positions are not renormalized due to tunnel coupling or electron-electron interactions. Such renormalizations would appear beyond the sequential tunneling regime or for reservoirs with energy resolved densities of states. Second, Eq. (12) shows that the occupation depends, quite generally, on voltage and temperature differences. Thus, $(-\frac{1}{\pi})\text{Im}\, G^r_{\sigma,\sigma}(E)$ will also depend on the applied fields, which will in turn affect the currents $I$ and $J$. This is a crucial fact that noninteracting models are not able to capture and turns out to be essential for a correct description of the nonlinear regime of transport. Our calculations thus have to be treated in a self-consistent fashion.

The self-consistent procedure can be simplified by noticing that due to the absence of Zeeman splittings the dot density should be nonmagnetic. Hence, the mean occupation in the dot $\langle n\rangle = \langle n_\sigma\rangle + \langle n_{\bar\sigma}\rangle$ reads

$$\langle n \rangle = \frac{2A}{1+A-B} \tag{13}$$

We have defined $A$ and $B$ as the integrals

$$A = \frac{1}{2\pi} \int dE \frac{\Gamma_L f_L(E) + \Gamma_R f_R(E)}{(E-\varepsilon_D)^2 + \Gamma^2/4} \tag{14}$$

$$B = \frac{1}{2\pi} \int dE \frac{\Gamma_L f_L(E) + \Gamma_R f_R(E)}{(E-\varepsilon_D-U)^2 + \Gamma^2/4} \tag{15}$$

In general, $\langle n \rangle$ takes values from 0 to 2 depending on the value of the energy level $\varepsilon_d$. The occupation is roughly constant between the Coulomb blockade peaks and significantly changes across resonances. Once the nonequilibrium occupation is determined for a given set of parameters, the currents $I$ and $J$ can be easily derived.

## 3. Heat Current

When a voltage is applied to an atomic-scale junction, experiments show an asymmetric rectification of the generated power [13]. We now investigate the heat flux distribution in a quantum dot using the theoretical model discussed above. Here, we consider that heat can flow due to voltage or to temperature biases. The former case is sketched in Fig. 1(a), where we explore a system with a symmetrically applied voltage bias and set $E_F = 0$. In the case of a temperature-driven dot as illustrated in Fig. 1(b), the temperature shift is applied to one of the electrodes: $\theta_L = \theta$ and $\theta_R = 0$ lead to positive temperature differences $\theta = T_L - T_R > 0$ while $\theta_L = 0$ and $\theta_R = \theta$ yield a negative temperature gradient.

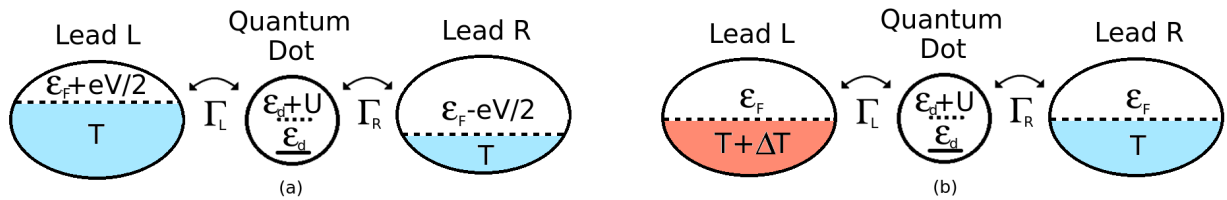

Fig. 1. Pictorial representations of (a) a quantum dot coupled to voltage biased terminals and (b) a quantum dot coupled to thermally biased reservoirs.

### 3.1. Voltage-driven case

Inserting Eq. (7) and the result of the retarded Green's function [Eq. (11)] in the expression for the heat current in Eq. (10), we calculate the dependence of $J$ as a function of the voltage bias (Fig. 2). In Fig. 2(a), we depict the heat current for several values of $\varepsilon_d$. At low bias (see the inset), $J$ depends linearly on $V$ (Peltier effect). Remarkably, this linear regime disappears quickly, allowing Joule and higher order effects to dominate (see the main figure). Interestingly, the heat current can be positive or negative at low bias (reversible heat), showing a nontrivial zero for finite $V$ when $\varepsilon_d = -3U/4$. This vanishing power is, in fact, more general: it occurs for any value of $\varepsilon_d$. Additionally, $J$ exhibits an asymmetric behavior for any value, expecting the particle-hole symmetry point $\varepsilon_d = -U/2$ but for different positions of $V$. As a consequence, a power asymmetry between $V$ and $-V$ arises. This asymmetry is apparent for large values of $\varepsilon_d$ far from the resonances. Our results in Fig. 2(a) show that there exists invariance when we simultaneously change $\varepsilon_d \to -\varepsilon_d - U$ and $V \to -V$.

The heat current changes its curvature for finite $V$. This is related to the maxima and minima of the electrothermal conductance (differential Peltier coefficient) $M = dJ/dV$, shown in Fig. 2(b). These extremal points are related to the alignment of the dot resonance peaks with the leads' electrochemical energies, yielding extremum values when $\varepsilon_d = -V/2$ and $\varepsilon_d = -V/2 - U$. We clearly see in Fig. 2(b) a change of sign depending of the sign of $V$. However, the point when the change of sign occurs is not $V = 0$ due to the nontrivial zero of the heat current. Furthermore, our results demonstrate that for sufficiently large voltages the electrothermal conductance $M$ directly follows from the electrical current, $M(V \to \infty) = -I/2$, because at very large voltages the occupation factors are simply $f_L \to 1$ and $f_R \to 0$ and the energy flux becomes independent of voltage.

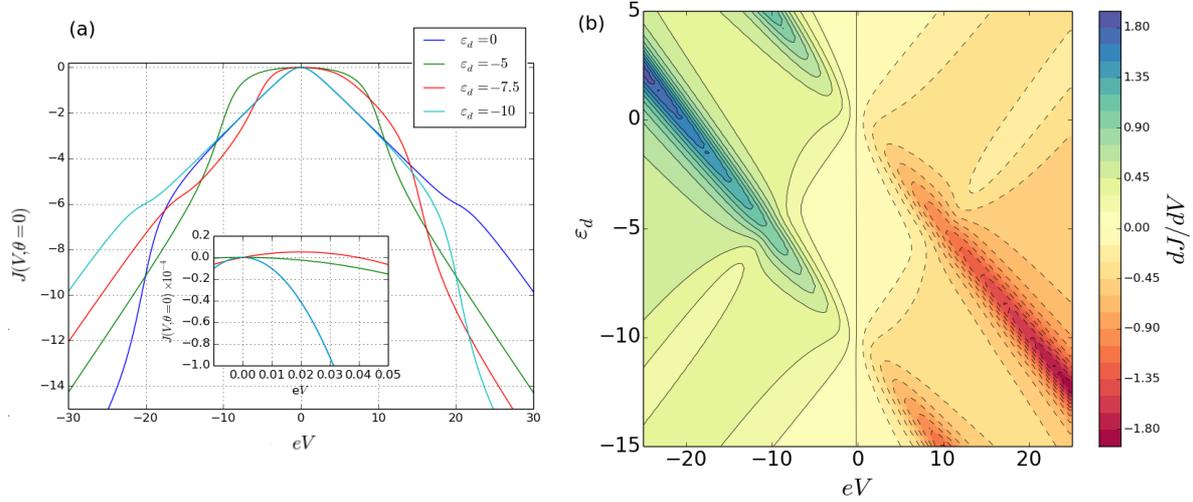

Fig. 2. (a) Heat current versus voltage difference for a single-level quantum dot in the Coulomb blockade regime and different level positions (gate voltages). In this case, the external temperature gradient is taken as $\theta = 0$. Inset: Detail of the heat current around the equilibrium point. (b) Differential electrothermal conductance $M$ as a function of the voltage bias and level position. All energies are set in terms of the broadening $\Gamma = \Gamma_L + \Gamma_R$ with $\Gamma_L = \Gamma_R$, charging energy $U$=10 $\Gamma$ and background temperature $T$=0.1 $\Gamma$.

### 3.2. Temperature-driven case

In Fig. 3(a), we show the heat current with an increasing temperature difference. $J$ is a monotonic function of $\theta$ for all values of the energy level. As in the voltage-driven case, the linear behavior here predicted by the Fourier law is quickly surpassed by higher order terms as the temperature shift exceeds values larger than the base temperature. The linear response behavior is better preserved when $\varepsilon_d = 0$ and $\varepsilon_d = -U$ (the two curves coincide) due to the alignment of the resonances with the electrochemical potentials. Furthermore, for increasingly larger values of $\theta$ the heat current attains a constant value. Physically, this can be understood from the fact that the occupation factor for thermally excited electrons becomes almost constant around the dot resonances for very high temperatures.

The saturation effect is related to the properties of the differential thermal conductance $K = dJ/d\theta$, which decays as $K \propto 1/\theta$ for large temperature biases. Figure 3(b) shows the thermal conductance as a function of both the temperature difference in the leads and the energy level $\varepsilon_d$. Clearly, $K$ is antisymmetric under reversal of $\theta$, meaning that the heat flux direction is determined by the hotter lead. We observe that near the resonances the thermal conductance reaches two maxima (or minima) before the power-law decay. In contrast, for energy levels around the particle-hole symmetry point, $\varepsilon_d = -U/2$, $K$ only reaches one extremal value. Therefore, heat conduction driven by thermal gradients is quite sensitive to variations of an external gate potential.

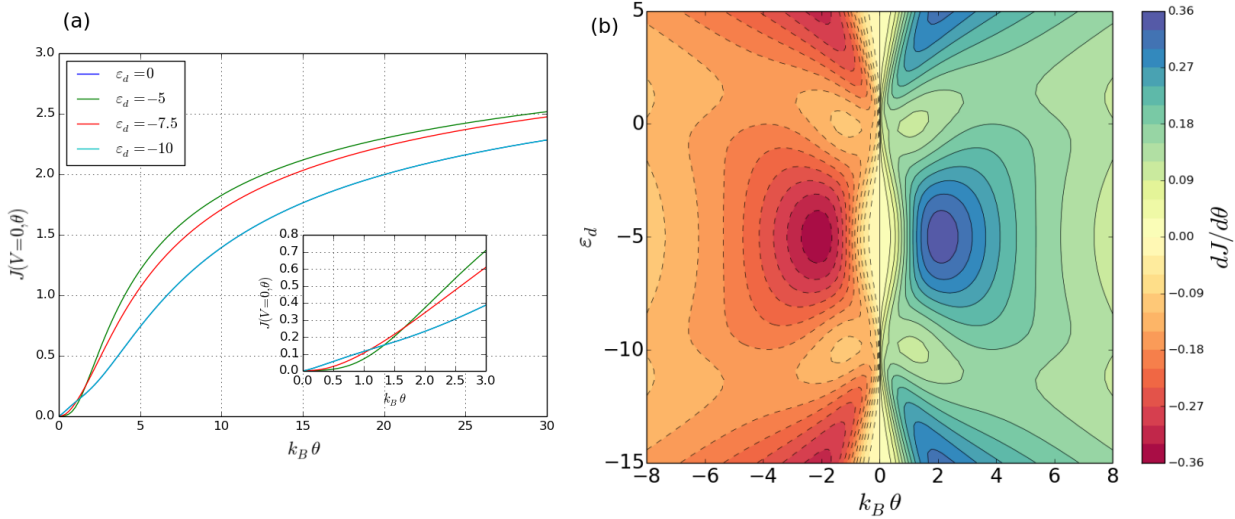

Fig. 3. (a) Heat current as a function of the temperature shift in absence of applied voltages and for different values of the gate voltage. The results for $\varepsilon_d = 0$ and $\varepsilon_d = -U$ are on top of each other. (b) Differential thermal conductance as a function of the temperature bias and level position. We use the same parameters as in Fig. 2.

## 4. Kelvin-Onsager relation

The Kelvin-Onsager relation establishes a reciprocal relation between the thermoelectric and electrothermal responses of a conductor out of equilibrium. In our system, the electrothermal conductance $M(V)$, which describes the amount of power generated in response to a voltage bias should be reciprocal to the thermoelectric conductance $L(\theta) = dI/d\theta$, which measures the current changes due to a shift of the temperature in the leads. Close to equilibrium, one has $M_0 = L_0 T$, where $M_0$ ($L_0$) are the electrothermal (thermoelectric) conductances at zero voltage and temperature differences ($V = \theta = 0$). This is a linear-response property with a universal character. As a consequence, it also applies to our Coulomb-blockaded dot. However, when nonlinear effects start to dominate, the relation is broken [25,26]. In Fig. 4 we illustrate this departure when the degree of nonlinearity increases with increasing $V$ or $\theta$. Clearly, in the limit of zero voltage or temperature bias the relation is exactly satisfied, but as the external fields increase, the deviation from the Kelvin-Onsager relation becomes apparent. Breakdown of the reciprocal relation occurs independently of $\varepsilon_d$, although is larger for values of the energy level that lie far from $\varepsilon_d = -U/2$ since at this point [Fig. 4(b)] the thermoelectric conductance exactly cancels due to particle-hole symmetry, independently of $\theta$.

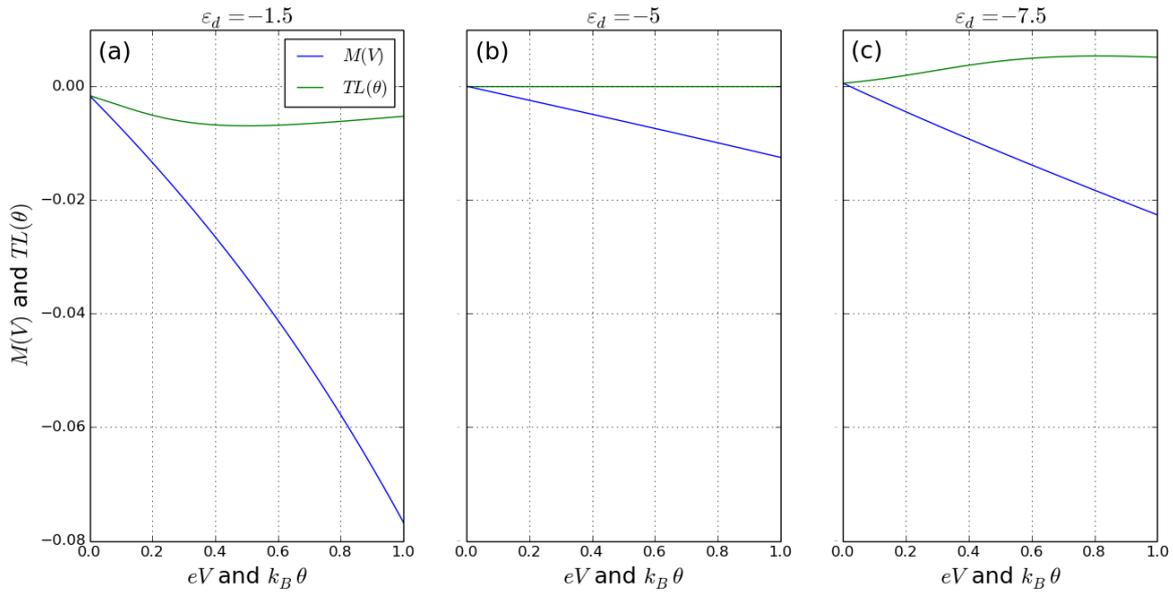

Fig. 4. Comparison between the differential electrothermal conductance *M* as a function of the voltage bias and the differential thermoelectric conductance *L* as a function of the temperature shift for three different gate voltages indicated above. We use the same parameters as in Figs. 2 and 3.

## 5. Conclusions

We have theoretically investigated the nonlinear power generated across a quantum dot in response to voltage or temperature shifts. The dot is largely dominated by charging effects yielding Coulomb blockade peaks in the local density of states. Based on our self-consistent calculations, we have shown that nonlinearities appear quickly in the heat current as voltage increases. The nonlinear Peltier coefficient shows a rich behavior as either the dc bias or the dot energy level is tuned. When the dot is subjected to a thermal gradient, the heat always flows from the hot reservoir to the cold one, leading to a distinctive antisymmetry for the differential thermal conductance. Finally, we have discussed breakdowns of the Kelvin-Onsager relation in the nonlinear regime of transport. Our work is useful to characterize the performance of small devices that aim at taking advantage of nonlinear effects in order to control heat dissipation at the nanoscale. Future works should consider the role of phonons and coherence beyond the sequential tunneling regime.